\documentstyle[preprint,aps,epsf,rotate]{revtex}

\begin{document}
\preprint{NSF-ITP 97-048}


\title{Comment on ``Duality relations for Potts correlation functions''}

\author{Jesper Lykke Jacobsen$^{1,2}$}

\address{$^1$Institute of Physics and Astronomy, University of Aarhus,
         DK-8000 Aarhus C, Denmark \\
         $^2$Somerville College and Department of Theoretical Physics,
	University of Oxford, 1 Keble Road, Oxford OX1 3NP, United Kingdom}

\date{\today}

\maketitle


\begin{abstract}
  In a recent paper by Wu \cite{Wu} the three-point correlation of the
  $q$-state Potts model on a planar graph was related to ratios of
  dual partition functions under fixed boundary conditions. It was
  claimed that the method employed could straightforwardly be applied
  to higher correlations as well; this is however not true. By
  explicitly considering the four-point correlation we demonstrate how
  the appearence of non-well-nested connectivities invalidates the method.

  \vspace{0.5cm}

\end{abstract}

Consider the $q$-state Potts model \cite{Potts} on a two-dimensional
planar graph $L$ having a free boundary. In accordance with Fig.~2 in
Ref.~\cite{Wu} we let $i$, $j$, $k$ and $l$ be four sites on the
boundary of $L$, following one another in a clockwise fashion, and we
define exterior dual spins $s_1$, $s_2$, $s_3$ and $s_4$ so that all
boundary spins between sites $l$ and $i$ of $L$ interact with a spin state
$s_1$, boundary spins between $i$ and $j$ interact with $s_2$, 
spins between $j$ and $k$ with $s_3$, and finally
spins between $k$ and $l$ with $s_4$. The
partition function with the four Potts spins $\sigma_i$, $\sigma_j$,
$\sigma_k$ and $\sigma_l$ fixed in definite states is called
$Z_{\sigma_i \sigma_j \sigma_k \sigma_l}$, and similarly the dual
partition function for fixed exterior dual spins is denoted by
$Z^{\ast}_{s_1 s_2 s_3 s_4}$.

Up to the $q$-fold permutation symmetry of the Potts spin labels there
exist 15 different boundary conditions for 
$Z_{\sigma_i \sigma_j \sigma_k \sigma_l}$
out of which we can form five combinations
\begin{eqnarray}
  Z_4    & = & Z_{1111}, \nonumber \\
  Z_3    & = & Z_{2111} + Z_{1211} + Z_{1121} + Z_{1112}, \nonumber \\
  Z_{\rm 2p} & = & Z_{2211} + Z_{2121} + Z_{2112}, \nonumber \\
  Z_{\rm p}    & = & Z_{1123} + Z_{1213} + Z_{1231} + Z_{2113} +
                                     Z_{2131} + Z_{2311}, \nonumber \\
  Z_0    & = & Z_{1234},
  \label{symm}
\end{eqnarray}
which are symmetric under permutations of the four sites $i$, $j$, $k$
and $l$. We introduce them here in order to simplify the notation in
subsequent equations.

Following Wu's strategy \cite{Wu} we should extract equations
relating the
$Z_{\sigma_i \sigma_j \sigma_k \sigma_l}$
and the
$Z^{\ast}_{s_1 s_2 s_3 s_4}$
by running through all possible ways of connecting the sites $i$, $j$,
$k$ and $l$ by auxiliary bonds and using the fundamental duality
relation, Wu's Eq.~(3). Each time a bond is added one of the
exterior dual spins $s$ is separated from its neighbours and thus allowed
to take a different value.

First, using the `empty' connection (i.e., introducing no auxiliary
bonds) we find the relation
\begin{equation}
  Z_4 + (q-1)[Z_3 + Z_{\rm 2p}] + (q-1)(q-2)Z_{\rm p} + (q-1)(q-2)(q-3)Z_0 =
     q C Z^{\ast}_4,
  \label{trivi}
\end{equation}
corresponding to Wu's Eq.~(19).

Next, consider adding a bond between sites $k$ and $l$. Summing over
the two `free' sites, $i$ and $j$, we obtain a reduction to the
two-point case:
\begin{eqnarray}
  Z_{11} & = & \sum_{i=1}^q \sum_{j=1}^q Z_{ij11} \nonumber \\
         & = & Z_{1111} + (q-1)[Z_{2111} + Z_{1211} + Z_{2211}] +
              (q-1)(q-2)Z_{2311}.
\end{eqnarray}
Among the exterior dual spins, $s_4$ has been separated from
$s_1 = s_2 = s_3$, so that $Z^{\ast}_{11} = Z^{\ast}_{1111}$ and
$Z^{\ast}_{12} = Z^{\ast}_{1112}$. Using the known result for
$Z_{11}$, Wu's Eq.~(12), along with the duality we find that
\begin{equation}
  Z_{1111} + (q-1)[Z_{2111} + Z_{1211} + Z_{2211}] +
             (q-1)(q-2)Z_{2311} =
  C[Z^{\ast}_{1111} + (q-1)Z^{\ast}_{1112}].
  \label{two-point}
\end{equation}
The equation is one out of a set of six obtained by connecting two of
the sites $i$, $j$, $k$ and $l$ with a bond. Further equations can be
found by considering $Z_{12}$ instead of $Z_{11}$, but they can be
shown to be linear combinations of Eq.~(\ref{trivi}) and the six
equations just obtained.

Another set of four equations can be found by letting one of the
sites $i$, $j$, $k$ and $l$ be `free' and connecting the remaining three.
This corresponds to a reduction to the three-point case. For example,
adding a bond between sites $j$ and $k$ and another between $k$ and
$l$, we find that
$Z_{111} = \sum_{i=1}^q Z_{i111} = Z_{1111} + (q-1)Z_{2111}$.
The exterior dual spins now satisfy $s_2 = s_1$, and from Wu's
Eq.~(25) for $Z_{111}$ and the duality we obtain
\begin{equation}
  Z_{1111} + (q-1)Z_{2111} = \frac{C}{q} \left \lbrace Z^{\ast}_{1111} +
    (q-1)[Z^{\ast}_{2211} + Z^{\ast}_{1121} + Z^{\ast}_{1112}] +
    (q-1)(q-2)Z^{\ast}_{1123} \right \rbrace.
  \label{three-point}
\end{equation}
Again, the equations found by reducing to Wu's expressions for
$Z_{123}$, $Z_{211}$, $Z_{121}$ or $Z_{112}$ do not contain any new
information.

The remaining equations can be found by connecting all of the sites
$i$, $j$, $k$ and $l$ with auxiliary bonds of strength $K$. One way of
doing this corresponds to Wu's Eq.~(24) and implies connecting all the
four sites to a common new point $n$ outside the boundary of
$L$. All of the exterior dual spins are now separated from one
another. Summing over $n$ we can express the partition function $\tilde{Z}$
of this system through the
$Z_{\sigma_i \sigma_j \sigma_k \sigma_l}$
and relate it by duality to its corresponding dual partition function
$\tilde{Z}^{\ast}$. The result is
\begin{eqnarray}
  Z_4 (e^{4K} &+& q-1) + (q-1) Z_3 (e^{3K}+e^K+q-2) +
    (q-1) Z_{\rm 2p} (2e^{2K}+q-2) \nonumber \\
  & + & (q-1)(q-2) Z_{\rm p} (e^{2K}+2e^K+q-3) + (q-1)(q-2)(q-3) Z_0 (4e^K+q-4)
    \nonumber \\
  & = & \frac{C}{q^2} (e^K-1)^4 \left \lbrace Z^{\ast}_4 e^{4K^{\ast}} +
    (q-1)[Z^{\ast}_3 + Z^{\ast}_{\rm 2p} - Z^{\ast}_{2121}] e^{2K^{\ast}} +
    (q-1)Z^{\ast}_{2121} \right. \nonumber \\
  & + & (q-1)(q-2) [Z^{\ast}_{\rm p} - Z^{\ast}_{1213} - Z^{\ast}_{2131}]
    e^{K^{\ast}} + (q-1)(q-2)[Z^{\ast}_{1213} + Z^{\ast}_{2131}]
    \nonumber \\
  & + & \left. (q-1)(q-2)(q-3) Z^{\ast}_0 \right \rbrace.
  \label{nontrivi}
\end{eqnarray}
The other way of connecting all four sites is to connect two of them
to a new point $m$ and the remaining two to another new point $n$. For
instance, connecting each of $i$ and $j$ to $m$ with a bond of
strength $K$, and similarly $k$ and $l$ to $n$, we find that $s_2$ and
$s_4$ are separated whilst $s_3 = s_1$. Summing over $m$ and $n$ as
before we arrive at
\begin{eqnarray}
  &&\Sigma_{11}^2 \left \lbrace Z_4 + (q-1)Z_{2211} \right \rbrace +
     \Sigma_{11} \Sigma_{21} \left \lbrace (q-1) Z_3 +
     (q-1)(q-2)[Z_{1123} + Z_{2311}] \right \rbrace
    \label{pairwise} \\
  & + & \Sigma_{21}^2 \left \lbrace (q-1) [Z_{2112} + Z_{2121}] +
        (q-1)(q-2)[Z_{\rm p} - Z_{1123} - Z_{2311}] +
        (q-1)(q-2)(q-3) Z_0 \right \rbrace \nonumber \\
  & = & \frac{C}{q} (e^K-1)^4 \left \lbrace  Z^{\ast}_4 e^{4K^{\ast}} +
        (q-1) [Z^{\ast}_{1211} + Z^{\ast}_{1112}] e^{2K^{\ast}} +
        (q-1) Z^{\ast}_{2121} + (q-1)(q-2) Z^{\ast}_{1213} \right \rbrace,
        \nonumber
\end{eqnarray}
where we have defined $\Sigma_{11} = e^{2K} + (q-1)$ and
$\Sigma_{21} = 2e^K + (q-2)$. A similar equation can be obtained by
connecting $i$ and $l$ to $m$, and $j$ and $k$ to $n$.

Exactly at this point we get into trouble. For the 15 unknown quantities
appearing on the right-hand side of Eq.~(\ref{symm}) we have obtained 14
independent equations, namely Eq.~(\ref{trivi}), six equations of the type 
(\ref{two-point}), four equations of the type (\ref{three-point}),
Eq.~(\ref{nontrivi}), and two equations of the type (\ref{pairwise}).
Clearly, what is missing is a third equation of the type
(\ref{pairwise}) in which $i$ and $k$ are connected to $m$, and $j$
and $l$ are connected to $n$. But the auxiliary bonds making such a
connection would necessarily intersect, thus violating the planarity of
the graph $L$. In other words, it would be impossible to define the
four exterior dual spins.

This problem has to do with the connectedness \cite{Blote} of the four
points. Generally, for an $n$-point correlation the number of
partition functions
$Z_{\sigma_1 \sigma_2 \ldots \sigma_n}$
with fixed values of the $n$ boundary spins equals the number of ways
$\tilde{c}_n$ in which $n$ points can be interconnected, as is easily seen by
interpreting $\sigma_i = \sigma_j$ as a connection between sites $i$
and $j$. But the number of equations obtainable using the method of Wu
is only equal to the number $c_n$ of well-nested $n$-point
connectivities \cite{Blote}, which is in general less than
$\tilde{c}_n$. For $n = 1,2,3,4,5,\ldots$ we have $\tilde{c}_n =
1,2,5,15,51,\ldots$, whilst $c_n = 1,2,5,14,42,\ldots$.

In an attempt to obtain more equations one could imagine introducing
connections between spins in the infinite face, thus converting any spin
of valence four or higher into interconnected spins of valence three
\cite{private}. In the case of the four-point function this would mean
discarding Eq.~(\ref{nontrivi}) and in addition to the two equations of
type (\ref{pairwise}) considering two new equations obtained from these
by connecting spins $m$ and $n$ with a fifth auxiliary bond. The first of
these, corresponding to Eq.~(\ref{pairwise}), looks like
\begin{eqnarray}
  & & Z_4 (\Sigma_{11}^2 + \Sigma_4) +
      (q-1) Z_3 (\Sigma_{11} \Sigma_{21} + \Sigma_3) +
      (q-1) Z_{2211} (\Sigma_{11}^2 + \Sigma_{\rm 2p}) \nonumber \\
  &+& (q-1)(q-2) [Z_{1123} + Z_{2311}] (\Sigma_{11} \Sigma_{21} + \Sigma_{\rm p}) +
      (q-1) [Z_{2121} + Z_{2112}] (\Sigma_{21}^2 + \Sigma_{\rm 2p}) \nonumber \\
  &+& (q-1)(q-2) [Z_{\rm p} - Z_{1123} - Z_{2311}] (\Sigma_{21}^2 + \Sigma_{\rm p}) +
      (q-1)(q-2)(q-3) Z_0 (\Sigma_{21}^2 + \Sigma_0) \nonumber \\
  &=& \frac{C}{q^3} (e^K - 1)^5 \left \lbrace Z_4^{\ast} e^{5K^{\ast}} +
      (q-1) [Z^{\ast}_{1211} + Z^{\ast}_{1112}] e^{3K^{\ast}}
      \right. \nonumber \\
  &+& (q-1) [Z^{\ast}_{2111} + Z^{\ast}_{1121} + Z^{\ast}_{2211} +
      Z^{\ast}_{2112}] e^{2K^{\ast}} + (q-1) Z^{\ast}_{2121} e^{K^{\ast}}
      \nonumber \\
  &+& \left. (q-1)(q-2)[Z^{\ast}_{\rm p} - Z^{\ast}_{2131}] e^{K^{\ast}} +
      (q-1)(q-2) Z^{\ast}_{2131} + (q-1)(q-2)(q-3) Z^{\ast}_0
      \right \rbrace,
  \label{trivalent}
\end{eqnarray}
where we have defined $\Sigma_4 = (e^K - 1)(e^{4K} + q - 1)$,
$\Sigma_3 = (e^K - 1)(e^{3K} + e^K + q - 2)$,
$\Sigma_{\rm 2p} = (e^K - 1)(2 e^{2K} + q - 2)$,
$\Sigma_{\rm p} = (e^K - 1)(e^{2K} + 2 e^K + q - 3)$, and
$\Sigma_0 = (e^K - 1)(4 e^K + q - 4)$.
Now, subtracting the sum of Eqs.~(\ref{trivalent}) from the sum of
Eqs.~(\ref{pairwise}) we regain, after some tedious algebra, $2 (e^K - 1)$
times Eq.~(\ref{nontrivi}). But, on the other hand, subtracting the
difference of Eqs.~(\ref{trivalent}) from the difference of
Eqs.~(\ref{pairwise}) we find the result {\em zero}, i.e., out of the four
equations where all the four spins are connected to auxiliary bonds there
are still only three that are linearly independent.

Although the method thus fails to isolate the
$Z_{\sigma_i \sigma_j \sigma_k \sigma_l}$,
which would enable us to construct the generalised correlation
function $P_4(\sigma,\sigma',\sigma'',\sigma''')$ as defined in Wu's
Eq.~(1), we could still hope to obtain the ordinary correlation
function $\Gamma_4(\sigma_i,\sigma_j,\sigma_k,\sigma_l)$ (see Wu's
Eq.~(2)). Evidently, to do so it suffices to isolate
$Z_4 \equiv Z_{1111}$, which should be possible by finding linear
combinations of the above set of equations for which the left-hand
sides are symmetric in the sense of Eq.~(\ref{symm}).

Clearly, Eqs.~(\ref{trivi}) and (\ref{nontrivi}) are already in this
form. The sum of the six equations of type (\ref{two-point}) yields
\begin{eqnarray}
  6 Z_4 + 3(q-1)Z_3 + 2(q-1)Z_{\rm 2p} &+& (q-1)(q-2)Z_{\rm p} = \nonumber \\
  && C \left \lbrace 6 Z^{\ast}_4 + (q-1)[Z^{\ast}_3 + Z^{\ast}_{\rm 2p} -
                       Z^{\ast}_{2121}] \right \rbrace,
\end{eqnarray}
while the four equations of type (\ref{three-point}) sum up to
\begin{eqnarray}
  4 Z_4 &+& (q-1)Z_3 = \nonumber \\
  && \frac{C}{q} \left \lbrace 4 Z^{\ast}_4 +
    (q-1)(q-2)[Z^{\ast}_{\rm p} - Z^{\ast}_{1213} - Z^{\ast}_{2131}] +
    (q-1)[2 Z^{\ast}_3 + 2(Z^{\ast}_{\rm 2p} - Z^{\ast}_{2121})] \right \rbrace.
\end{eqnarray}
As before, employing the two-point reduction $Z_{12}$ or the
three-point reductions $Z_{123}$, $Z_{211}$, $Z_{121}$ and $Z_{112}$
merely provides us with linear combinations of what we already know.

But again, the two equations of the type (\ref{pairwise}) lead us
into trouble. When taking their sum it can easily be seen that the
left-hand side cannot be written in terms of the symmetrical
combinations (\ref{symm}). To symmetrise we would again need the
non-well-nested equation, which however leaves the exterior dual spins
undefined.

In conclusion we have shown that the method of Wu breaks down for the
four-point correlation, and in fact for all higher correlations as
well. This is reminiscent of the fact that in conformal field theory
it is necessary to consider non-planar, as well as planar, monodromies
in order that the four- and higher-point correlation functions be well
determined \cite{monodromy}.
The failure to relate the four-point function to generalised
surface tensions is especially regrettable in so far as such a relation might
prove useful in calculating the energy-energy correlation of the Potts
model.

Useful discussions with J.~L.~Cardy and F.~Y.~Wu are gratefully acknowledged.
This research was supported in part by the Engineering and Physical Sciences
Research Council under Grant GR/J78327, and by the National Science Foundation
under Grant PHY94-07194.


\end{document}